# Towards a Taxonomy of Performance Evaluation of Commercial Cloud Services

Zheng Li
School of CS
NICTA and ANU
Canberra, Australia
Zheng.Li@nicta.com.au

Liam O'Brien
CSIRO eResearch
CSIRO and ANU
Canberra, Australia
Liam.OBrien@csiro.au

Rainbow Cai
School of CS
NICTA and ANU
Canberra, Australia
Rainbow.Cai@nicta.com.au

He Zhang
School of CSE
NICTA and UNSW
Sydney, Australia
He.Zhang@nicta.com.au

*Abstract* — **Cloud Computing, as one of the most promising computing paradigms, has become increasingly accepted in industry. Numerous commercial providers have started to supply public Cloud services, and corresponding performance evaluation is then inevitably required for Cloud provider selection or cost-benefit analysis. Unfortunately, inaccurate and confusing evaluation implementations can be often seen in the context of commercial Cloud Computing, which could severely interfere and spoil evaluation-related comprehension and communication. This paper introduces a taxonomy to help profile and standardize the details of performance evaluation of commercial Cloud services. Through a systematic literature review, we constructed the taxonomy along two dimensions by arranging the atomic elements of Cloud-related performance evaluation. As such, this proposed taxonomy can be employed both to analyze existing evaluation practices through decomposition into elements and to design new experiments through composing elements for evaluating performance of commercial Cloud services. Moreover, through smooth expansion, we can continually adapt this taxonomy to the more general area of evaluation of Cloud Computing.**

I. INTRODUCTION

Among all the emerging computing paradigms, Cloud Computing has been viewed as the most promising one in the current computing industry [1]. There are an increasing number of providers offering Cloud infrastructures and services with different terminology, definitions, and goals [2]. Considering customers have little knowledge and control over the precise nature of commercial Cloud services even in the "locked down" environment [3], performance evaluation of those services would be crucial for many purposes ranging from cost-benefit analysis for Cloud Computing adoption to decision making for Cloud provider selection. In fact, the effort on performance evaluation of commercial Cloud services emerged as soon as those services were published [9, 19]. Therefore, it is not necessary to study performance of Cloud services and the corresponding evaluation approaches from scratch. Existing evaluation experiences especially the typical experimental methods can be summarized and reused. To get familiar with the existing approaches to performance evaluation of commercial Cloud services, we have identified 46 relevant studies from 2006 to 2010 through a systematic literature review. When investigating these studies, unfortunately, we found many issues that could obstruct comprehending and spoil drawing lessons from the existing evaluation work. The issues are mainly threefold:

- Non-standardized terminology. For example, (1) the authors in [5] confused Cloud Provider and Cloud Service, and thus readers have to distinguish them by context; (2) Performance Variability has been depicted as Performance Stability [6], Performance Homogeneity [6] and Performance Fluctuation [7].
- Correct but imprecise analysis. For example, different sizes of Message Passing Interface (MPI) messages were used to evaluate the communication-throughput variability of EC2 [8]. However, this type of experiment is particularly for evaluating scalability, although the poor scalability in this case was due to back-end variation.
- Incorrect analysis. For example, the failure-number related metrics were used to measure Availability of commercial Cloud services [9], whereas those metrics are actually for measuring Reliability that is driven by the number of failures [10].

Consequently, it is necessary to clarify the possibly-confusing concepts and inaccurately-used terminology existing in the current Cloud performance evaluation work. Considering a well founded taxonomy is significantly beneficial to corresponding research in any field of study [4], we have established a novel taxonomy of performance evaluation of commercial Cloud services. This paper specifies the taxonomy including brief introductions to the establishment process and some application scenarios.

A particular characteristic of our work is to identify the atomic elements of performance evaluation in the context of Cloud Computing. In detail, the taxonomy is constructed along two dimensions: Performance Feature and Experiment. Moreover, the Performance Feature dimension is further split into *Physical Property* and *Capacity* parts, while the Experiment dimension is split into *Environmental Scene* and *Operational Scene* parts, as illustrated in Figure 1. Therefore, in addition to the general benefits from taxonomy like providing common terminology, the taxonomy proposed in this paper can be further employed both to analyze existing evaluation practices through decomposition into elements and to design new experiments through composing elements for evaluating commercial Cloud services. Moreover, we may continually adapt this taxonomy to the more general area of Cloud Computing through smooth expansion [4].

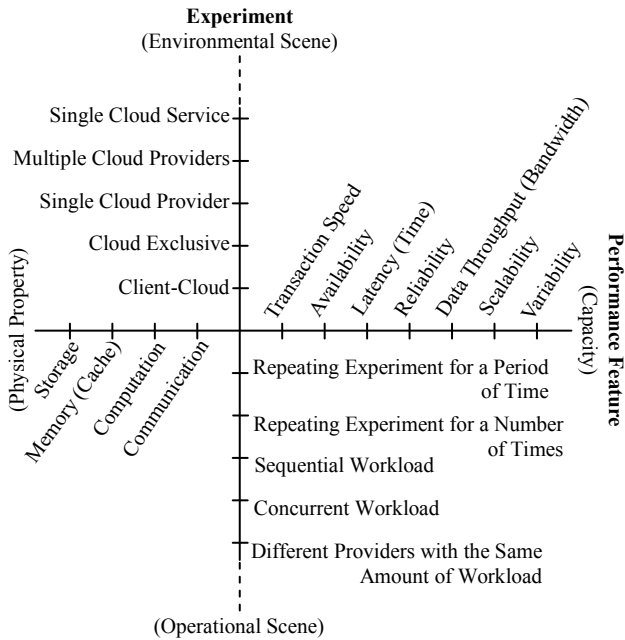

Figure 1. Taxonomy of performance evaluation of Cloud services.

Note that there is no arrowed direction in the above coordinates-like illustration. In other words, Figure 1 does not imply any linear relationship among those atomic elements. Furthermore, due to the space limit, Figure 1 does not completely show all the experimental scenes (see Section IV) in its vertical axis. The dashed lines in the Experiment dimension axis indicate that there are more elements in both *Environmental Scene* part and *Operational Scene* part.

The remainder of the paper is organized as follows. Section II briefly introduces the procedure used to build up the taxonomy particularly for performance evaluation of commercial Cloud services. Section III and IV respectively specify the taxonomy's two dimensions and their elements in detail. Section V suggests two application scenarios of this proposed taxonomy. Conclusions and some future work are discussed in Section VI.

## II. METHODOLOGY OF ESTABLISHING THE TAXONOMY

Like the way most taxonomies were established, this proposed taxonomy is also generated through a regression manner – abstraction of knowledge from the existing primary work. In particular, we have focused on the real evaluation practices in the context of commercial Cloud Computing. To efficiently achieve the corresponding knowledge, we adopt the systematic literature review (SLR) to investigate the existing approaches to evaluation of commercial Cloud services. As the main methodology applied for Evidence-Based Software Engineering (EBSE) [20], SLR has been widely accepted as a standard and systematic approach to investigation of specific research questions by identifying, assessing, and analyzing published primary studies. According to the guidelines of SLR [21], we used mainly three steps to unfold this research:

- Identify research questions and prepare the SLR.
- Select relevant primary studies and extract data.
- Analyze the extracted data and report the result.

### A. Research Questions

In the first step, we only highlight the research questions instead of specifying the complete preparation work for SLR. Intuitively, one performance evaluation instance is composed of the evaluated performance feature and the corresponding experiment, while an experiment includes experimental environment and operations. Therefore, we have drawn three research questions as listed in Table I. The answers to these questions could be initially qualitative descriptions, and then be refined into atomic elements of performance evaluation to establish the taxonomy.

TABLE I. RESEARCH QUESTIONS

| ID | Research Question | Motivation |
|---|---|---|
| Q1 | What performance features of commercial Cloud services have been evaluated? | To identify the atomic elements of evaluated performance features. |
| Q2 | In the evaluation experiments, what environments/resources have been employed? | To identify the atomic elements of experimental environments for evaluation. |
| Q3 | In the evaluation experiments, what operations have been manipulated? | To identify the atomic elements of experimental operations for evaluation. |

### B. Selecting Primary Studies and Extracting Data

To answer the above research questions, relevant primary studies should be selected, and by using rational and efficient search strategies. Firstly, the time window of candidate publications was set. Considering that the term "Cloud Computing" started to gain popularity in 2006 [22], we can focus on the literature published from the beginning of 2006 onwards. And also considering the possible delay of publishing, we have restricted our search to the period between *January 1st, 2006* and *December 31st, 2010*. Secondly, we prescribed a set of criteria for strictly deciding whether or not a primary study should be reviewed. The selection criteria are not listed in this paper due to the limit of space. In particular, this work focused only on the commercial Cloud services to make our effort closer to industry's needs.

Overall, we have identified 46 relevant primary studies covering six commercial Cloud providers from a set of popular digital publication databases (study list for reference: http://www.mendeley.com/groups/1104801/slr4cloud/papers/). The qualitative data can then be extracted for further analysis by answering the pre-defined research questions.

### C. Data Analysis

The data analysis here is mainly to identify the common description or terminology among all the qualitative answers. The identified descriptions and terms have been abstracted and finally composed into a two-dimensionally structured taxonomy, as illustrated in Figure 1. Both taxonomy dimensions are further divided into two parts, and the component elements are specified in the following two sections respectively.

III. PERFORMANCE FEATURES OF COMMERCIAL CLOUD SERVICES FOR EVALUATION

In the Performance Feature dimension of this taxonomy, we are not to define new terms but to rationalize the terms already in use. In practice, one evaluated performance feature is usually represented by a combination of a physical property of Cloud services and its capacity, for example Communication Latency, or Storage Reliability. Therefore, we split the Performance Feature dimension into two parts: *Physical Property* part and *Capacity* part. In particular, although both Scalability and Variability have been viewed as Cloud service's aspects independent of Performance [11, 12], they are inevitably reflected by the change of value of performance features. Considering their close relationship with performance, and for the convenience of discussion, here we also regard Scalability and Variability as two elements in the *Capacity* part but distinguished from the other capacities. Thus, all the elements in the Performance Feature dimension can be organized as shown in Figure 2, and each of them is specified in the following subsections.

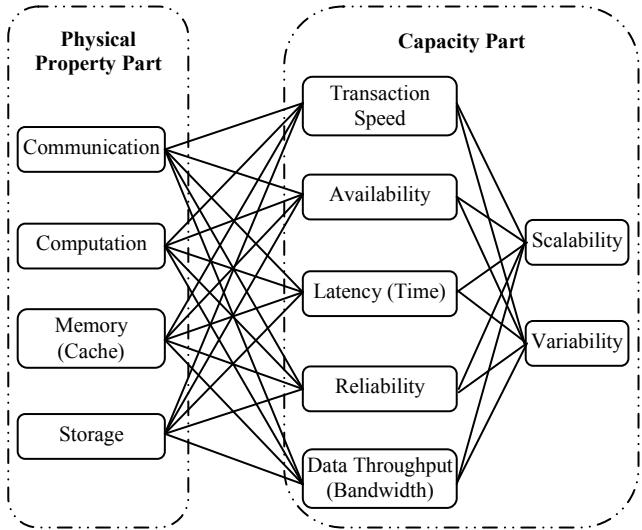

Figure 2. Performance features of Cloud services for evaluation.

A. *Physical Property Part*

1) *Communication:* Communication refers to the data/message transfer between client and the Cloud, different Cloud services, or different service instances. Strictly, Communication is not an internal physical property of Cloud services, whereas it will inevitably appear when employing Cloud services through Internet/network.

2) *Computation:* Computation refers to the computing-intensive data/job processing in the Cloud. Typically, the virtual CPU could be directly evaluated to reflect the Computation of a commercial service, if the service exists as virtual machine (VM) instances.

3) *Memory (Cache):* Here we treat Memory and Cache as a unified physical property of Cloud services, because both memory and cache are intended for fast access to temporarily saved data that can be achieved from slow-accessed hard drive storage. Note that we only concern the inherent memory/cache of Cloud services [17] rather than the man-made cache system by using Cloud resources [13].

4) *Storage:* Contrasted with Memory (Cache), Storage of Cloud services can be used to permanently store users' data, until the data are removed or the services are suspended intentionally. Storage can be either the main functionality or partial supplied respectively by different Cloud services, for example Amazon S3 vs. EC2.

B. *Capacity Part*

1) *Transaction Speed:* A transaction may be an independent job or pre-defined operation related to a physical property of Cloud services, for example one float calculation of CPU or disk read/write operation. In contrast with the term Throughput that is generally used for data transfer, we adopt Speed to describe how fast transactions can be processed.

2) *Availability:* Availability is driven by the time lost [10], which describes the probability a system works in functioning condition during a specific period of time. Given an interval, Availability of a commercial Cloud service can be calculated as a ratio of the uptime of the Cloud service to the total time of the interval, usually in a yearly basis. For example, 99.95% availability of EC2 is claimed by Amazon, which indicates 4.3 hours of non-scheduled downtime per year [14].

3) *Latency (Time):* Latency is mainly related to the measure of time delay for a particular job. In the literature, we can find that the definition of latency varies depending on different contexts or perspectives. Here we give Latency the broadest meaning to describe all the time-related capacities of a commercial Cloud service. As for the detailed contexts and perspectives, we can make further analysis by distinguishing different time windows.

4) *Reliability:* Reliability is driven by the number of failures [10], which describes the probability a system can properly perform its intended function during a specific period of time. Note the tricky difference between the definitions of Availability and Reliability: a system works in a functioning condition does NOT necessarily indicate that it performs its intended function. Although the value of Reliability may be also expressed as a percentage with a time basis, the typical metrics reflecting reliability are mainly related to accounting individual failures, for example Mean Time Between Failure (MTBF) and Failure Rate (FR) [18]. Consequently, we can draw the distinction between evaluations for Availability and Reliability by detecting the metrics whether involving the failed time or involving the failure number.

5) *Data Throughput (Bandwidth):* In tradition, Throughput is the term used mainly for measuring data communication. Here we consider Throughput as a general concept that describes an amount of data processed in a particular period of time (from input to output) by any physical property of Cloud services. Moreover, we also replaced another widely-used term Bandwidth with Throughput, though they are subtly different concepts. In fact, the ideal bandwidth of a physical property can be viewed as one of the inherent characteristics of a Cloud service, which has to be reflected though evaluating the actual throughput. In other words, the actual throughput is equal to the effective bandwidth of a particular physical property in real evaluation experiments.

*6) Scalability:* Similar to the property Latency, Scalability may also be variously defined within different contexts or from different perspectives [19, 11]. From the perspective of changing resource (with a certain amount of workload), Scalability can be distinguished between *Horizontal Scalability* and *Vertical Scalability*. The former means the ability of employing more resources, while the latter stands for the ability of increasing the power of resources [16]. From the perspective of changing workload (with unchanged resource), Scalability refers to the ability of a system to deal with the gradually increasing amount of work in a graceful manner. Imagine the naming convention of *Horizontal* and *Vertical Scalability* is in coordinates, we name the Scalability from the second perspective of changing workload as *Original Scalability*.

*7) Variability:* In general, Variability describes the state of spread of a set of data. In this paper, we use Variability to indicate the extent of fluctuation in values of an individual performance feature of a commercial Cloud service. In the existing literature, performance variability of Cloud services has been further specified as either performance stability or performance homogeneity [6]. Performance stability concerns the time changing of using Cloud services, which reflects the performance fluctuation over time when employing the same Cloud resource; Performance homogeneity concerns the environmental changing of using Cloud services, which reflects the performance fluctuation when either service instances or clients are not identical.

To sum up, through dividing every performance feature of Cloud services into two parts, we have identified four physical properties and seven capacity elements in the Performance Feature dimension of this taxonomy. Following the linkages between the elements in the *Physical Property* part and the *Capacity* part, we can exhaustively reveal different performance features of a particular commercial Cloud service.

## IV. EXPERIMENTAL SCENES IN PERFORMANCE EVALUATION OF COMMERCIAL CLOUD SERVICES

In contrast with the Performance Feature dimension, we define a set of new terms/phrases for the elements in the Experiment dimension of this taxonomy. In fact, we already borrowed the term "scene" from drama to name this dimension. In the context of drama, a scene is an individual segment of a plot in a story, and usually settled in a single location. By analogy, here we use "scene" to represent an atomic unit for constructing a complete experiment for evaluating commercial Cloud services. Moreover, we further distinguish between environmental scenes and operational scenes. As the names suggest, environmental scenes concern experimental resources that can together profile an evaluation environment, whilst operational scenes reflect experimental processes with human interference. In detail, we have identified ten environmental scenes and 15 operational scenes that are specified in the following subsections respectively, which can be organized as an experimental scene tree, as shown in Figure 3.

### A. Environmental Scenes

As mentioned previously, environmental scenes indicate static descriptions that are used to specify required experimental resources. In particular, non-Cloud resources collaborated with Cloud services are all regarded as Client.

*1) Single Cloud Provider vs. Multiple Cloud Providers:* When it comes to the distinction between these two scenes, it should be clarified that the context here is one single experiment job. In other words, a primary study covering multiple Cloud providers does not imply that it has the evaluation scene *Multiple Cloud Providers*, unless an experiment job is designed to be finished by collaboration between different providers' services. For example, suppose a primary study evaluates two providers' Cloud services respectively for performance contrast, we should consider the *Single Cloud Provider* scene for this study, because the experiment jobs for each service are independent and can be separated into different evaluation experiments.

*2) Single Cloud Service vs. Multiple Cloud Services:* Similarly, we define *Multiple Cloud Services* scene as the situation that there is a single experiment job involving two or more different Cloud services, otherwise *Single Cloud Service* scene. Note that only the same provider's services in an experiment job should be considered when recognizing these two scenes.

*3) Single Service Instance vs. Multiple Service Instances:* The concept "service instance" is only applicable to the Cloud services that can supply service in units, like virtual machine (VM) instances of EC2. Meanwhile, it does not make sense to count service instances supplied by different Cloud services. Therefore, we only concern applicable and single Cloud service when recognizing these two scenes.

*4) Single Instance Type vs. Multiple Instance Types:* The identification of these two scenes is not based on the reviewed studies, but on such a consideration that although we have tried to exhaustively reveal the environmental scenes in the reviewed primary studies, new experimental scene(s) could still exist. For example, theoretically, it is possible to assign different functional roles to different types of VM instances to finish a single experiment job, which will bring an environmental scene *Multiple Instance Types*. However, we have not found such experiment jobs in the reviewed literature. In other words, we recognize that all the reviewed evaluation works only concerned the *Single Instance Type* scene.

*5) Cloud Exclusive vs. Client-Cloud:* This pair of scenes is relatively straightforward. *Cloud Exclusive* refers to that an experiment job that is completely performed by using Cloud resources, while *Client-Cloud* indicates that resources from both inside and outside of the Cloud are used to finish an experiment job. Note that the *Cloud Exclusive* scene does not require Cloud resources coming from the same provider.

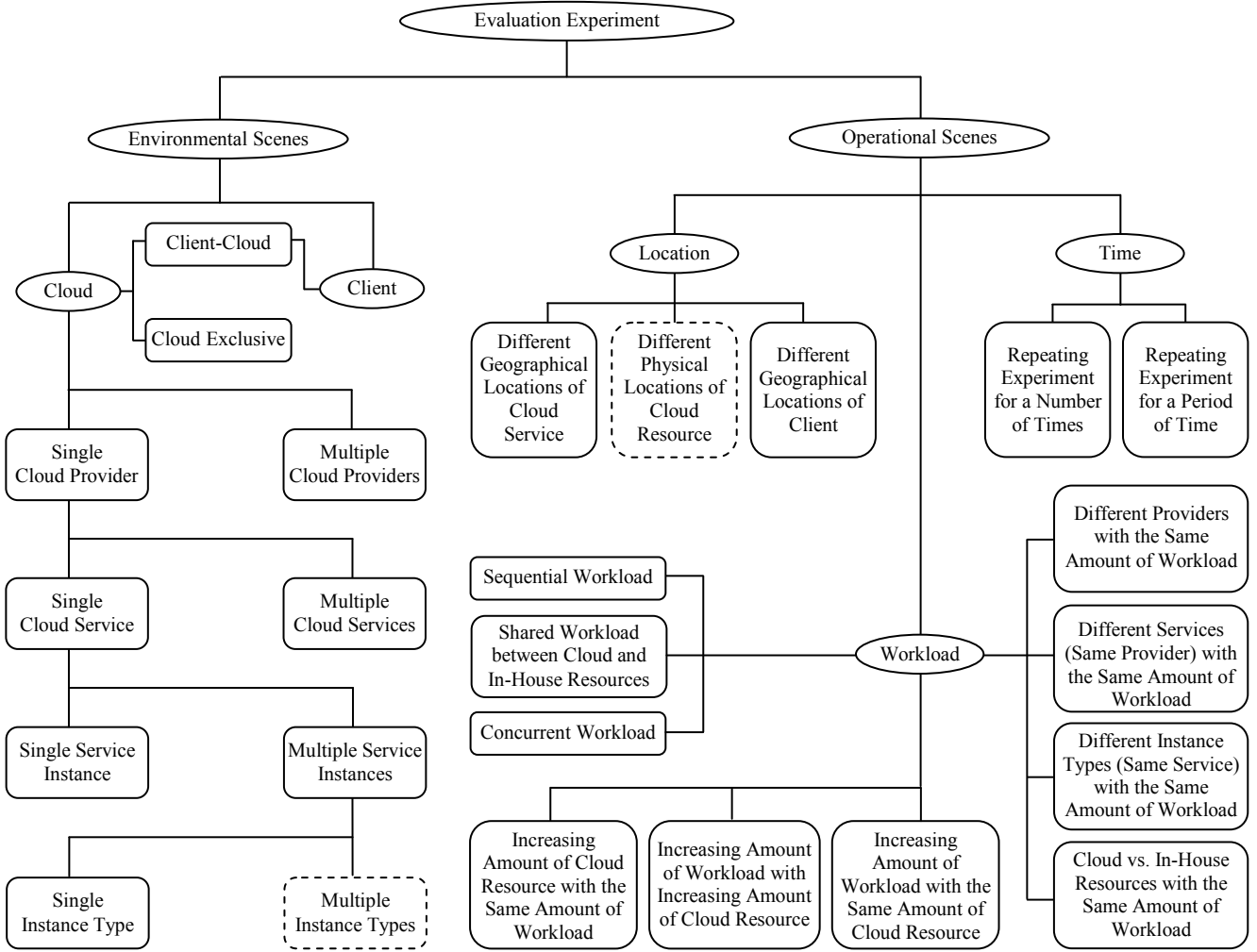

Figure 3. Experimental scene tree of performance evaluation of commercial Cloud services.

*B. Operational Scenes*

In contrast with environmental scenes, operational scenes indicate dynamic manipulations that usually imply repeating an individual experiment job under different circumstances. In particular, the operational scenes are further grouped according to their keywords: time, location, and workload. Note that non-Cloud resources in operational scenes are all named as In-House resource.

*1) Repeating Experiment for a Period of Time:* This scene is to repeat an individual experiment job during a particular period of time. Such a scene is particularly suitable for experiment jobs that can finish in tiny time segments. Through serializing a set of same experiment jobs, we can implement evaluation more easily in a longer time.

*2) Repeating Experiment for a Number of Times:* This scene is to repeat an individual experiment job for a particular number of times. In contrast with the previous scene, this one is particularly suitable for long-time experiment jobs. Naturally, frequency control is not only easier but also more accurate than period setting for long-time running experiments.

*3) Different Physical Locations of Cloud Resource:* Although not common, this scene particularly indicates the same type of virtual machines (VMs) running with un-virtualized differences. The un-virtualized difference here refers not only to the difference in underlying hardware like different model of real CPU, but also to the difference between VMs sharing or not sharing underlying hardware.

*4) Different Geographical Locations of Cloud Service:* The geographical location of Cloud service depends on the place where the corresponding Cloud data center is hosted. For reasons such as disaster discovery or performance improvement, Cloud providers may deploy many data centers in different locations around the world. Therefore, this sense is mainly used to investigate the effect of service-side locations on performance of Cloud services.

*5) Different Geographical Locations of Client:* In contrast with the previous scene, the clients interacted with Cloud services may also be located in different places over the world. The aim of this scene is then to investigate the effect of client-side locations on performance of Cloud services.

*6) Sequential Workload:* Sequential workload usually consists of a series of successive transactions. Each transaction request will not be issued until the previous one is submitted. Generally, this scene is used to investigate the serial processing performance of Cloud services.

*7) Concurrent Workload:* Concurrent workload indicates that multiple transaction requests are simultaneously submitted to Cloud service, and a set of batches of simultaneous requests can be issued successively. In contrast with the previous scene, this one is used to investigate the parallel processing performance of Cloud services.

*8) Shared Workload between Cloud and In-House Resources:* This scene describes that an experiment job is completed through collaboration between In-House and Cloud resources. The collaboration refers to that the workload in the experiment job is handled by amount sharing or functionality sharing.

*9) Increasing Amount of Cloud Resource with the Same Amount of Workload:* This scene appears when using different amount of resource to repeat dealing with a particular amount of workload. Note that we can still draw this scene from practice even if decreasing amount of Cloud resource in the repeating experiments. In this case, the experimental results can be analyzed along the opposite direction of the real experimental activities.

*10) Increasing Amount of Workload with the Same Amount of Cloud Resource:* Unlike the previous scene, this one is to use a particular amount of resource to deal with different amount of workload respectively. Note that here the entire capability of the given resource is emphasized. We do not further identify whether or not the entire resource's capability is saturated by the given workload.

*11) Increasing Amount of Workload with Increasing Amount of Cloud Resource:* If changing amount of both workload and resource in an individual experiment job, this scene can be abstracted, and only abstracted from the scenario that workload increases along with resource.

*12) Different Providers with the Same Amount of Workload:* For the reason of business competition, different Cloud providers may offer functionally similar services. Therefore, this scene can be typically drawn when performing evaluation for comparison between Cloud services supplied by different providers.

*13) Different Services (Same Provider) with the Same Amount of Workload:* To satisfy different requirements, one Cloud provider may supply different while comparable services. For example, Amazon EC2, EBS and S3 all provide data storage service, but with different costs and benefits [13]. Therefore, this scene appears when evaluating several candidate services while considering performance as one of the trade-off factors.

*14) Different Instance Types (Same Service) with the Same Amount of Workload:* This scene typically exists in the evaluation for comparing different types of instances of a particular Cloud service. Similar to the "instance" concept discussed in aforementioned environmental scenes, we also only consider applicable and single Cloud service when mentioning instance types.

*15) Cloud vs. In-House Resources with the Same Amount of Workload:* This scene is to run the same experiment job by using Cloud resource and In-House resource respectively. The common intention of performing extra experiment with In-House resource is to build baseline for evaluating Cloud resources [15]. Moreover, this scene also exists in the performance evaluations for comparison between Cloud Computing and other computing paradigms.

Overall, through the experimental scene tree, as shown in Figure 3, we may easily locate or enumerate individual environmental and operational scenes when evaluating commercial Cloud services. In particular, the rounded rectangle with dashed line represents the scene either uncontrollable (*Different Physical Locations of Cloud Resource*) or unemployed yet (*Multiple Instance Types*).

V. APPLICATIONS OF THE TAXONOMY

As mentioned in the motivation of establishing this taxonomy, we can in turn use the established taxonomy to facilitate analyzing the existing practices of performance evaluation of commercial Cloud services. Moreover, generic experiences of performance evaluation can be summarized through these analyses within a standard context of terminology. Benefitting from the summarized experiences, we can further design new experiments for evaluating emerging Cloud services or complementing current performance evaluation work.

*A. Analyzing Existing Practices of Performance Evaluation for Commercial Cloud Services*

To avoid duplication of the primary studies used for building this taxonomy, we employ a relatively new publication [23] to demonstrate the analysis. Meanwhile, the demonstration can also be viewed as a partial validation of the taxonomy proposed in this paper.

The analysis result is listed in Table II. In the study introduced by [23], five tasks were settled on for evaluating performance of Cloud services supplied by four commercial providers. Through mapping the authors' work with this taxonomy, we can find that the evaluation details are able to be abstracted into and interpreted by a set of performance properties with corresponding experimental scenes clearly.

If marking those analysis-identified elements in the aforementioned two-dimensional taxonomy space (see Figure 1), each task is then represented as a set of points. By visibly illustrating the abovementioned analysis, we can conveniently distinguish between similar evaluation experiments, and then identify particular characteristics of different evaluations. For example, in the case of particularly evaluating Transaction Speed of Computation, the un-overlapped points indicate that the Single-Machine Compute task requires *Single Service Instance* and *Sequential Workload*, while the Multi-Machine HPL task requires *Multiple Service Instances* with *Concurrent Workload* and concerns Horizontal Scalability by using the operational scene *Increasing Amount of Resource with the Same Amount of Workload*.

Through such a way of investigation, more evaluation characteristics, some of which are listed in the next subsection, can be identified and interpreted by using this taxonomy.

TABLE II.  ANALYSIS OF PERFORMANCE EVALUATION OF COMMERCIAL CLOUD SERVICES IN PRACTICE [23]

| Evaluation Task | Performance Feature | | Experiment | |
|---|---|---|---|---|
| | *Physical Property* | *Capacity* | *Environmental Scenes* | *Operational Scenes* |
| Single-Machine Compute Performance | Computation | Transaction Speed also (Vertical) Scalability | Cloud Exclusive Single Cloud Provider Single Cloud Service Single Service Instance | Sequential Workload Repeating Experiment for a Number of Times Different Providers with the Same Workload Different Instance Types (Same Service) with the Same Workload |
| Single-Machine I/O Performance | Memory(Cache) Storage | Data Throughput Transaction Speed (Original) Scalability also (Vertical) Scalability | Cloud Exclusive Single Cloud Provider Single Cloud Service Single Service Instance | Sequential Workload Repeating Experiment for a Number of Times Increasing Amount of Workload with the Same Amount of Resource Different Providers with the Same Workload Different Instance Types (Same Service) with the Same Workload Cloud vs. In-House Resources with the Same Workload |
| Multi-Machine HPL Performance | Computation | Transaction Speed also (Vertical) Scalability (Horizontal) Scalability | Cloud Exclusive Single Cloud Provider Single Cloud Service Multiple Service Instances Single Instance Type | Concurrent Workload Repeating Experiment for a Number of Times Increasing Amount of Resource with the Same Amount of Workload Different Providers with the Same Workload Different Instance Types (Same Service) with the Same Workload |
| Multi-Machine HPCC Performance | Computation Communication Memory(Cache) Storage | Transaction Speed Data Throughput Latency also (Vertical) Scalability (Horizontal) Scalability | Cloud Exclusive Single Cloud Provider Single Cloud Service Single Service Instance Multiple Service Instances Single Instance Type | Concurrent Workload Repeating Experiment for a Number of Times Increasing Amount of Resource with the Same Amount of Workload Different Providers with the Same Workload Different Instance Types (Same Service) with the Same Workload Cloud vs. In-House Resources with the Same Workload |
| Single-Machine Performance Stability | Computation Memory(Cache) | Data Throughput Variability also (Original) Scalability | Cloud Exclusive Single Cloud Provider Single Cloud Service Single Service Instance | Sequential Workload Repeating Experiment for a Number of Times Increasing Amount of Workload with the Same Amount of Resource Different Providers with the Same Workload |

*B. Designing Experiments for Performance Evaluation for Commercial Cloud Services*

In fact, the performance evaluation analysis and design are bilateral activities of using this proposed taxonomy. As such, aforementioned illustration can also facilitate designing evaluation experiments by pointing necessary elements in the two-dimensional space. Meanwhile, the experimental experiences gathered through evaluation analysis can be reused for future practices of performance evaluation. Considering the straightforward process of designing evaluation experiments with illustration, here we only list some typical evaluation experiences instead of completely designing an evaluation case for Cloud services.

According to our investigation, many unclear and even confusing descriptions can be found in the work of evaluating Scalability and Variability of commercial Cloud services. Hence, we particularly specify the summarized evaluation experiences for these two capacities.

*1) Scalability:*
- The operational scene *Different Instance Types (Same Service) with the Same Amount of Workload* is particularly suitable for evaluating Vertical Scalability of Cloud services.
- The operational scene *Increasing Amount of Cloud Resource with the Same Amount of Workload* is particularly suitable for evaluating Horizontal Scalability of Cloud services.
- The operational scene *Increasing Amount of Workload with the Same Amount of Cloud Resource* is particularly suitable for evaluating Original Scalability of Cloud services.

*2) Variability:*
- The time-related operational scenes *Repeating Experiment for a Number of Times* and *Repeating Experiment for a Period of Time* are particularly suitable for evaluating Variability (in terms of Stability) of Cloud services.
- Two of the location-related operational scenes *Different Physical Locations of Cloud Resource* and *Different Geographical Locations of Cloud Service* are particularly suitable for evaluating Variability (in terms of Cloud-side Homogeneity) of Cloud services.
- The location-related operational scene *Different Geographical Locations of Client* is particularly suitable for evaluating Variability (in terms of client-side Homogeneity) of Cloud services.

## VI. CONCLUSIONS AND FUTURE WORK

Since Cloud Computing is still maturing, internal changes may happen at times within current Cloud services, and new commercial providers may gradually appear. Hence, performance evaluation of different commercial Cloud services could be required all the time. As such, it is worthwhile to summarize existing performance evaluation experiences rather than the evaluation results, and then reuse them in the continual evaluation practices. To facilitate drawing lessons from the existing efforts, we established a novel taxonomy of performance evaluation of commercial Cloud services. Benefitting from the evaluation elements supplied by this taxonomy, we can conveniently deal with performance evaluation of Cloud services through a divide-and-conquer approach. Existing evaluation work can be analyzed through decomposition into elements for experience summarization, while new experiments can be designed through composing elements to satisfy future evaluation requirements.

Our future work related to this research is twofold. Firstly, we plan to build an "apple-to-apple" evaluation model for commercial Cloud services based on this taxonomy. As previously mentioned, different commercial providers offer different Cloud services with different standards and specifications [2]. It is impossible to implement absolutely fair and comparable performance evaluations for all the Cloud services. However, we can establish an abstract evaluation model by using standardized terminology, and then give unified guidelines for detailed evaluation practices. Secondly, we plan to enlarge this taxonomy's applicable area. In fact, commercial Cloud services also have other features in addition to the Performance, such as Security and Economics [24]. Through smooth expansion, we can make this taxonomy adapt to the more general area of evaluation of Cloud Computing.